\documentclass[useAMS,twocolumn]{mn2e}
\usepackage{amssymb}
\usepackage{graphicx}

\usepackage[dvips]{color}

\title[The role of initial and boundary conditions]
 {On the role of initial and boundary conditions in numerical simulations of accretion flows}
\author[Bu et al.]
{De-Fu Bu$^{1}$\thanks{E-mail:dfbu@shao.ac.cn}, Feng
Yuan$^{1}$\thanks{E-mail:fyuan@shao.ac.cn}, Maochun Wu$^{1,2}$
, and Jorge Cuadra$^{1,3}$\\
$^{1}$Key Laboratory for Research in Galaxies and Cosmology,
Shanghai Astronomical Observatory,\\ Chinese Academy of
Sciences, 80 Nandan Road, Shanghai 200030, China\\
$^{2}$Graduate School of the Chinese Academy of Sciences, Beijing
100039, China\\
$^{3}$Departamento de Astronomia y Astrofisica, Pontificia
Universidad Catolica de Chile, Chile \\}

\date{Accepted . Received ; in original form}

\begin{document}

\maketitle

\begin{abstract}
We study the effects of initial and boundary conditions, taking
two-dimensional hydrodynamical numerical simulations of hot
accretion flow as an example. The initial conditions considered
include a rotating torus, a solution expanded from the
one-dimensional global solution of hot accretion flows, injected gas
with various angular momentum distributions, and the gas from a
large-scale numerical simulation. Special attention is paid to the
radial profiles of the mass accretion rate and density. Both can be
described by a power-law function, $\dot{M}\propto r^s$ and
$\rho\propto r^{-p}$. We find that if the angular momentum is not
very low, the value of $s$ is not sensitive to the initial condition
and lies within a narrow range, $0.47\la s \la 0.55$. However, the
value of $p$ is more sensitive to the initial condition and lies in
the range $0.48\la p \la 0.8$. The diversity of the density profile
is because different initial conditions give different radial
profiles of radial velocity due to the different angular momentum of
the initial conditions. When the angular momentum of the accretion
flow is very low, the inflow rate is constant with radius. Taking
the torus model as an example, we have also investigated the effects
of inner and outer boundary conditions by considering the widely
adopted ``outflow'' boundary condition and the ``mass flux
conservation'' condition. We find that the results are not sensitive
to these two boundary conditions.
\end{abstract}


\section{INTRODUCTION}
Hot accretion flows, such as advection-dominated accretion flows
(ADAFs; Ichimaru 1977; Rees et al. 1982; Narayan \& Yi 1994; 1995;
Abramowicz et al. 1995; see Narayan, Mahadevan \& Quataert 1998 and
Kato, Fukue \& Mineshige 1998 for reviews) are interesting because
they are likely operating in low-luminosity active galactic nuclei
(AGNs), including the supermassive black hole in our Galactic
center,  and hard and quiescent states of black hole X-ray binaries
(see Narayan 2005; Yuan 2007; Narayan \& McClintock 2008; Ho 2008;
Yuan 2011 for reviews). Because of its widespread
applications, the properties of hot accretion flows have been
investigated intensively in the last decade. Hot accretion flows are
originally proposed and studied analytically by
vertically-integrated one-dimensional method (e.g., Narayan 1994).
In the analytical works, the mass accretion rate is usually assumed
to be constant with radius. Under this assumption, the radial
profile of density satisfies $\rho (r) \propto r^{-3/2}$.

Later on, intensive numerical simulation works have been performed
to investigate the hot accretion flows. One of the most important
finding is that the mass inflow rate is not a constant of radius;
rather, it decreases inward. Correspondingly, the density profile
flattens. The density profile obtained by the simulations is $\rho
\propto r^{-p}$, with $0.5\la p \la 1$. This is true for both
hydrodynamic (HD) and magnetohydrodynamic (MHD) simulations (e.g.,
Stone, Pringle \& Begelman 1999, hereafter SPB99; Igumenshchev \&
Abramowicz 1999, 2000, hereafter IA99, IA00; Stone \& Pringle 2001;
Hawley, Balbus \& Stone 2001; Machida, Matsumoto \& Mineshige 2001;
Hawley \& Balbus 2002; Igumenshchev, Narayan \& Abramowicz 2003;
Pen, Matzener \& Wong 2003; De Villiers, Hawley \& Krolik 2003; Yuan
\& Bu 2010; Pang, et al. 2011; McKinney, Tchekhovskoy \& Blandford
2012; Li, Ostriker \& Sunyaev 2013; see Yuan, Wu \& Bu 2012 for a
review). One model proposed to explain the above result is ADIOS
(adiabatic inflow-outflow solution). In this scenario, the inward
decrease of inflow rate is because of mass lost via outflow
(Blandford \& Begelman 1999, 2004; Begelman 2012).

By analyzing the HD and MHD numerical simulation data, Yuan, Bu \&
Wu (2012, hereafter YBW2012; see also Li, Ostriker \& Sunyaev 2013)
show that outflow must exist, thus confirming the ADIOS scenario.
Moreover, they investigated the physical mechanism of producing
outflows. In the HD case, they find that the temperature of outflow
is systematically higher than that of inflow (Fig. 8 in YBW2012).
This suggests that the outflow is driven by the buoyancy which
arises because of the convective instability of the HD accretion
flow. In the case of MHD flow, they find that the specific angular
momentum of outflow is close to the Keplerian angular momentum;
while that of inflow is much lower (Fig. 11 in YBW2012). This
suggests that it is the centrifugal force associated with the
magnetic field that drives the outflow (Blandford \& Payne 1982).

The radial profile of mass accretion rate (or density) is important
because of the following reasons. The profile of mass accretion rate
determines the emitted spectrum and other radiative features of an
accretion flow (e.g., Quataert \& Narayan 1999; Yuan, Quataert \&
Narayan 2003). With the increasing power of telescopes, the gas
properties (density, temperature) at the Bondi radius of galaxies
can been observed directly in some sources, such as Sgr A* and
NGC~3115. Using the observational data and Bondi accretion theory
(Bondi 1952), we can obtain the mass accretion rate at Bondi radius.
Then depending on the radial profile of mass accretion rate (or more
exactly the density), the radiation of the accretion flow is
completely different. In cosmology simulations and other simulations
studying the accretion flow on relatively large scale (e.g., Cuadra
et al. 2005; 2006; 2008; Springel et al. 2005; Booth \& Schaye 2009;
Power et al. 2011; Hobbs et al. 2012), we can at most resolve the
Bondi radius and determine the Bondi accretion rate. Then the mass
accretion rate profile can determine the evolution of black hole
mass and spin. The radial profile of mass accretion rate is also
important to determine the feedback efficiency of an AGN. It is well
known that feedback from AGN can strongly influence the physical
properties of their vicinity, their host galaxies, and even of the
intergalactic material of galaxy clusters to which they belong
(e.g., Proga, Stone \& Kallman 2000; Proga 2007; Kurosawa \& Proga
2009; Ciotti \& Ostriker 1997; 2001; 2007; Ciotti, Ostriker \& Proga
2009; Novak, Ostriker \& Ciotti 2011). Both the radiation and mass
outflow from a AGN can influence its surroundings. The radiation
from AGN can heat or cool the gas surrounding it, and the efficiency
of heating or cooling depends on the spectrum and the conditions of
the gas. The outflows can directly blow away the gas surrounding an
AGN. If the Bondi accretion rate can be estimated from observations,
the mass accretion rate profile within the Bondi radius can then
determine how much radiation and outflow can be produced and further
influence the environment of AGNs.

In this paper, we will study how the properties (especially radial
profiles of mass accretion rate and density) of hot accretion flows
depend on the initial and boundary conditions. Our motivations are
as follows. In the realistic universe, the environment of accretion
flows is likely very complex and differ from each other.
Correspondingly, the initial conditions of accretion flows, such as
their radial velocity, temperature, and angular momentum, should be
very different in different sources. In the literature, torus and
injection models are usually adopted as the initial condition of
simulations. 
YBW2012 studied several HD models. In their model A, the initial
condition is a rotating torus. In model C, they inject gas at the
outer boundary. Although the description and strength of viscosity
are same in different models, the properties of the accretion flows
are found to be quite different. First, in model A, the flow is
nearly symmetric to the equatorial plane, with inflow being close to
the equatorial plane ($\theta=70^{\circ}-90^{\circ}$ in a spherical
coordinate) and outflow close to the surface of the disk
($\theta=50^{\circ}$ and $140^{\circ}$). In model C, the flow is not
symmetric to the equatorial plane, and large-scale bulk motions
dominate the flow. Second, in model C, the radial velocity of inflow
close to the equatorial plane can be described as $v_r \propto
r^{-0.5}$; but in model A, the radial velocity scaling with radius
is much steeper. The difference between the results obtained by
different initial settings motivates us to systematically study the
influence of the initial conditions on the properties of hot
accretion flows.

In the present work, we focus on HD simulation. In reality, people
now believe that it is the MHD turbulence associated with the
magneto-rotational instability (MRI; Balbus \& Hawley 1991; 1998)
that is responsible for the angular momentum transfer in accretion
flows. Therefore, MHD simulation is more realistic. Our
considerations of adopting HD simulation are as follows. First, MHD
simulation is obviously more expensive; and the results depend on
the initial configuration and strength of the magnetic field.
Second, in many astrophysical simulations such as large-scale
cosmological ones, magnetic field is not considered and HD
simulations are still the dominant ones. It is thus of practical
usefulness to consider HD simulations. Third, we expect that if the
initial and boundary conditions are important to determine some
properties of HD accretion flow, it should also be important for MHD
flows. Moreover, by comparing all relevant HD and MHD numerical
simulations, Yuan, Wu \& Bu (2012) found that the radial profiles of
inflow rate obtained by various HD and MHD simulations are very
similar. In other words, it seems that the result does not depend on
the presence or absence of magnetic field, the strength and
configuration of the magnetic field, and the dimension of
calculation (two or three dimension). Such an apparently surprising
result was predicted and explained in Begelman (2012). This implies
that maybe the results obtained in the present study can be applied
to MHD case as well.

The structure of the paper is as follows. In \S 2, we describe the
basic equations, various initial and boundary conditions adopted and
our numerical method. The results of simulations will be given in \S
3. We summarize our results in \S 4.

\section{Numerical method}

\subsection{Equations}

In spherical coordinates $(r, \theta, \phi)$, we solve the following
hydrodynamical equations describing accretion:
\begin{equation}
\frac{d\rho}{dt}+\rho\nabla\cdot \mathbf{v}=0,\label{cont}
\end{equation}
\begin{equation}
\rho\frac{d\mathbf{v}}{dt}=-\nabla p-\rho\nabla
\psi+\nabla\cdot\mathbf{T}, \label{rmon}
\end{equation}
\begin{equation}
\rho\frac{d(e/\rho)}{dt}=-p\nabla\cdot\mathbf{v}+\mathbf{T}^2/\mu,
\label{rmon}
\end{equation}
Here, $\rho$, $p$, $\mathbf{v}$, $\psi$, $e$ and $\mathbf{T}$ are
density, pressure, velocity , gravitational potential, internal
energy and anomalous stress tensor, respectively. $d/dt(\equiv
\partial / \partial t+ \mathbf{v} \cdot \nabla)$ denotes the Lagrangian
time derivative. We adopt an equation of state of ideal gas
$p=(\gamma -1)e$, and set $\gamma =5/3$.

We use the stress tensor $\mathbf {T}$ to mimic the shear stress,
which is in reality magnetic stress associated with MHD turbulence
driven by MRI. Following SPB99, we assume that the only non-zero components of
$\mathbf {T}$ are the azimuthal components,
\begin{equation}
  T_{r\phi} = \mu r \frac{\partial}{\partial r}
    \left( \frac{v_{\phi}}{r} \right),
\end{equation}
\begin{equation}
  T_{\theta\phi} = \frac{\mu \sin \theta}{r} \frac{\partial}{\partial
  \theta} \left( \frac{v_{\phi}}{\sin \theta} \right) .
\end{equation}
This is because the MRI is driven only by the shear associated with
orbital dynamics. Other components of the stress are much smaller
than the azimuthal components (Stone \& Pringle 2001). We adopt the
coefficient of shear viscosity $\mu=\nu\rho$. We assume $\nu\propto
r^{1/2}$, which is the usual ``$\alpha$'' description. In all
models, we set $\alpha$=0.01 to eliminate the discrepancy due to the
magnitude of viscosity coefficient.

We use pseudo-Newtonian potential to mimic the general relativistic
effects, $\psi=-GM/(r-2r_g)$, where M is the center black hole mass,
G is the gravitational constant and $r_g \equiv GM/c^2$ is the
gravitational radius. The self gravity of the gas is neglected.

\subsection{Models}

\begin{table*}
\footnotesize
\begin{center}
\caption{Models in this paper}
\begin{tabular}{ccccccccccccc} \\ \hline
Run  & Descriptions &$N_r \times N_\theta $& {$l (f_\phi^1)$} &  {$v_r (f_r^2)$} & $e(f_e^3)$  & {$t_f$} & $r_{out}$ & $s^4$ & $p^5$\\
\hline

A & torus      & 168$\times$  88 & 0.9-1.1 &  0.0 &     &4.5 & $800r_g$ & 0.49 & 0.53 \\
A1$^6$ & tours     & 168 $\times$  88 & 0.9-1.1 &  0.0 &     &4.5 & $800r_g$ & 0.5 & 0.54 \\
Ah1 & torus      & 252$\times132$  & 0.9-1.1 &  0.0 &     &4.5 & $800r_g$ & 0.49 & 0.52 \\
Ah2 & torus      & 336$\times$ 176 & 0.9-1.1 &  0.0 &     &4.5 & $800r_g$ & 0.49 & 0.52 \\
B & G-ADAF-S     & 200 $\times$  88 & 0.4-0.8 &  0.01-1 & 0.2   &10. & $600r_g$ & 0.55 & 0.8 \\
C1 & injection & 168 $\times$  88 & 0.95      &  0.1  & 0.2  &900. & $600r_g$ & 0.53 & 0.51\\
C2 & injection & 168 $\times$  88 & 0.55      &  0.1  & 0.2  &900. & $600r_g$ & 0.54 & 0.49\\
C3 & injection & 168 $\times$  88 & 0.55      &  0.01 & 0.2  &900. & $600r_g$ & 0.55 & 0.48\\
C4 & injection & 168 $\times$  88 & 0.25      &  0.1  & 0.2  &900. & $600r_g$ & 0.54 & 0.63\\
C5 & injection & 168 $\times$  88 & 0.1       &  0.1 & 2.0  &900. & $600r_g$ & 0 & 1.31\\
C6 & injection & 168 $\times$  88 & 0.1       &  0.1 & 4.0  &900. & $600r_g$ & 0 & 1.3\\
$D^\dagger$ & injection & 168 $\times$  88  & 0.6       &  0.01-0.1
&0.1  &600. & $600r_g$ & 0.47 & 0.65\\

\hline
\end{tabular}\\
1,$f_{\phi}$ is the ratio of angular momentum to the local Keplerian
angular momentum for the initial tours or injected gas. \\
2,$f_r$ is the ratio of radial velocity to the local Keplerian
velocity for the initial tours or injected gas. \\
3,$f_e$ is the ratio of the internal energy to the local
gravitational energy for the initial torus or the injected gas.\\
4, s is the power law index of the radial profile of inflow rate.\\
5, p is the power law index of the radial profile of density. \\
6, In model A1, at the outer radial boundary, we use mass-flux conservation boundary conditions. \\
 $^\dagger$ \textbf{The properties of the injected gas in model
D are more "realistic".}
\end{center}
\end{table*}
For our aim, we adopt four kinds of initial conditions. In all our
models, the inner boundaries locate at $r=2.7r_g$. In the angular
direction, the boundary conditions are set by symmetry at the poles.
In model A1, at the outer radial boundary, we use mass-flux
conservation boundary conditions (see Section 3.4); For the inner
radial boundary in model A1, we use outflow boundary conditions. In
all of other models, at the inner and outer radial boundaries, we
use outflow boundary conditions.

In model A series, the initial condition of our simulation is an
equilibrium torus with constant specific angular momentum given by
(Papaloizou \& Pringle 1984)

\begin{equation}
\frac{p}{\rho} = \frac{(\gamma-1)GM}{\gamma R_{0}} \left[
\frac{R_{0}}{r} - \frac{1}{2} \left( \frac{R_{0}}{r \sin \theta}
\right)^{2} - \frac{1}{2d} \right] .
\end{equation}
Here, $R_{0}$ is the radius of the center (density maximum) of the
torus, and $d$ is the distortion of the torus. We assume $R_{0}=200
r_g$ and $d=1.25$ in our models. The maximum density of the tours
$\rho_{max}=1.0$.  The specific angular momentum of the initial
tours equals to the Keplerian angular momentum at the torus center.
Initially, the torus is embedded in a low-density medium. The
density and pressure of the medium are $\rho_m=10^{-4}$ and $
p_m=\rho_m/r$, respectively.

In model B, we use the one-dimensional steady global ADAF solution
(G-ADAF-S; Yuan 1999) as our initial condition. Because this
solution is one-dimensional, we extend it into two-dimensions in the
following way. The values of ${\rho(r,0)}$, $ {p(r,0)}$, $
v_r(r,0)$, $l(r,0)$ are from the global solution of ADAF. The
density and pressure of flow exponentially decreases along $\theta$:
$\rho(r,\theta)={\rho(r,0)} \rm{exp(-z^2/2H^2)}$,
$p(r,\theta)={p(r,0)}\rm{exp(-z^2/2H^2)}$, where $z=r  |{\cos
\theta}|$. For simplicity, we set the radial velocity $
v_r(r,\theta)=v_r(r,0)$ and azimuthal velocity $
v_{\phi}(r,\theta)={{l(r,0)}/(r{\rm sin(\theta)})}$. Initially, the
$\theta$ component of the velocity is set to $0$. In Figure 1, we
plot the radial structure of these initial quantities.

\begin{figure}
\includegraphics[width=8.5cm]{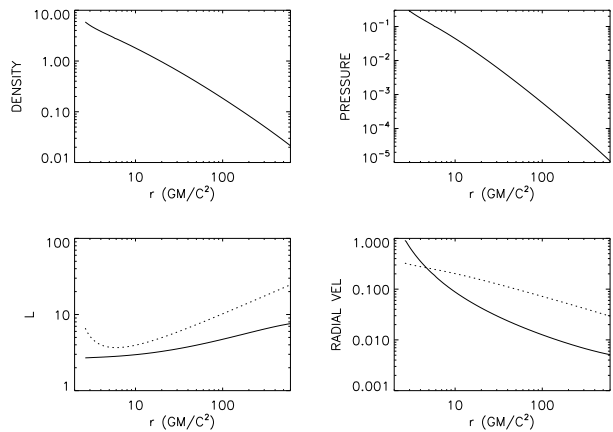}
\caption{One-dimensional global solution of hot accretion flows on
which the initial condition of Model B is based. The dotted line in
the angular momentum plot shows the Keplerian one; the dotted line
in the radial velocity plot shows the sound speed.}
\end{figure}

Model C series are ``injection'' type. The gas is continually injected
into the computational domain from the outer boundary. The $\theta$
component of the velocity of the injected gas is set to be zero. The
density of the injected gas has a Gaussian distribution about the
equator $\rho(r,\theta)=\rho_{max} \rm
{exp[-0.5(\theta-{\pi}/2)^2]}$, with $\rho_{max}=1$. In model C series,
we have assumed that the injected gas have different angular moment,
radial velocity and internal energy. The rotational and radial
velocities of the injected gas are $v_\phi =f_\phi v_k$ and $v_r=
{f_r}v_k$, respectively, with  $v_k$ being the Keplerian velocity at the
injection radius. The internal energy of the injected fluid $e=
{f_e}\rho\psi$. In models C1-C4, $f_e$ is set to 0.2, so the gas in
these models is marginally bound, i.e. the Bernoulli parameter
$Be < 0$. The detailed parameters are listed in Table 1.

In model D, the initial condition is adopted to be more
``realistic''. We inject gas in full-range of ${\theta}$. The
properties of the injected gas are taken from a parsec-scale,
realistic SPH simulation of the gas dynamics around Sgr A* (see
Cuadra et al 2006, 2008 for details).  The gas in this simulation
originates as stellar winds of young massive stars, whose orbital
and mass-loss properties are well determined from observations
(Paumard et al 2006; Martins et al 2007).  Shocks between different
stellar winds heat up the gas, creating a hot accretion flow. They
ran the simulation for $\sim1000\,$yr, which corresponds to several
orbital times for the stars that originate the gas. In order to
apply the 3 dimensional data from the simulation by Cuadra (2008) to
the current 2 dimensional simulation, we have to chose a symmetry
axis -- we take the vector normal to the plane on which most stellar
orbits lie as the rotation axis. We then average the data over the
azimuthal angle to obtain 2 dimension data. Then, the 2 dimensional
data was used as initial conditions in model D presented in this
work. Looking at these data, we find the average specific angular
momentum of the gas at the inner boundary of their simulation
($5000r_g$) is $l\approx 0.6 l_k$. The radial velocity has large
fluctuation, and is $\sim 0.01- 0.1 v_k$. The temperature is $\sim 7
\times 10^7 $ K. We use the method of least squares with equation
$log(y)=a+b*log(r)$ (y denotes the physical quantities such as
specific energy, angular momentum and radial velocity) to fit the
results of Cuadra et al. (2008). After we obtain the value of $a$
and $b$, we can  extrapolate the result of Cuadra et al. to obtain
the physical quantities used at the outer boundary ($600r_g$) in
model D.

Table 1 summarizes the properties of the models mentioned-above.
Column (2) gives a brief description of the models (initial state is
torus, global ADAF solution or injection from the outer boundary).
Column (3) shows the numerical resolution. Columns (4)-(6) show the
properties of the gas in the initial conditons. They are the
specific angular momentum compared to the local Keplerian angular
momentum, the radial velocity compared to the local Keplerian
velocity, the specific internal energy compared to the local
gravitational energy. Column (7) gives the final time at which we
stopped the simulation (in units of the Keplerian orbital time at
$r= 200$ ${r_g}$). Column (8) denotes the outer boundary. When the
simulations achieve steady state, the power-law index of the radial
profiles of mass inflow rate and density are given in columns (9)
and (10), respectively.

\begin{figure}
\includegraphics[width=8.5cm]{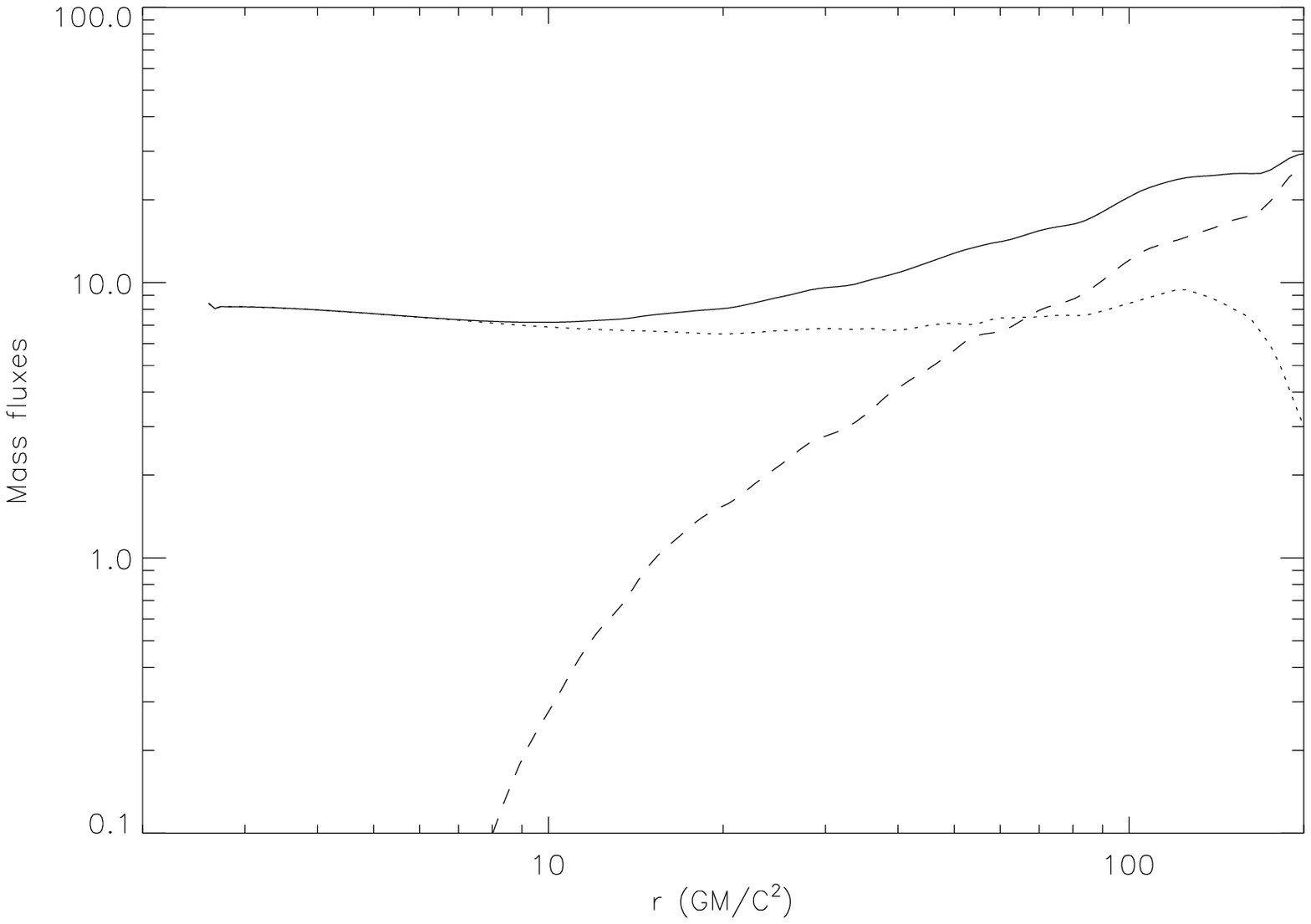}
\includegraphics[width=8.5cm]{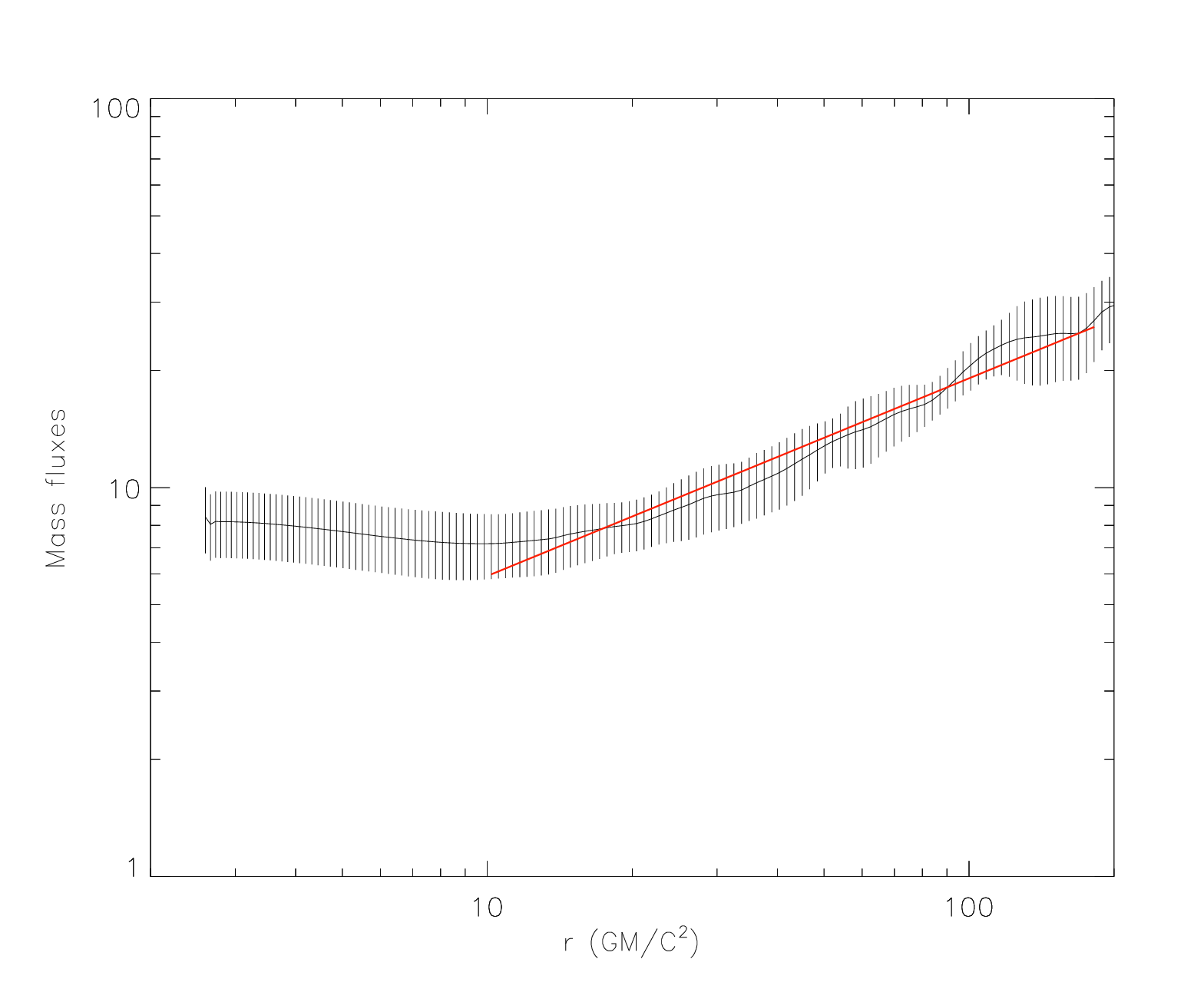}
\caption{Top: The radial profile of the time-averaged and angle
integrated mass inflow rate $\dot{M}_{\rm in}$ (solid line), outflow
rate $\dot{M}_{\rm out}$ (dashed line), and the net rate
$\dot{M}_{\rm acc}$ (dotted line) in model A. They are defined in
equations (\ref{inflowrate}), (\ref{outflowrate}), and
(\ref{netrate}), respectively. Bottom: Power law fit to the inflow
rate. The red line is our fit to the time-averaged inflow rate. The
black line is the time-averaged inflow rate. In this panel, we also
show the standard deviation of the time-averaged inflow rate.}
\end{figure}

\subsection{Numerical method}

We use the ZEUS-2D code (Stone \& Norman 1992a,1992b) to solve
equations (1)-(3). We adopt non-uniform grid in the radial direction $(\bigtriangleup r)_{i+1} /
(\bigtriangleup r)_{i} = 1.037$. Similarly, we adopt non-uniform
angular zones with $(\bigtriangleup \theta)_{j+1} / (\bigtriangleup
\theta)_{j} = 0.9826$ for $0 \leq \theta \leq \pi/2$  and
$(\bigtriangleup \theta)_{j+1} / (\bigtriangleup \theta)_{j} =
1.0177$ for $\pi/2 \leq \theta \leq \pi$.

\section{RESULTS}

Following SPB99, we define the mass inflow and outflow rates, $\dot
{M}_{in}$ and $\dot {M}_{out}$, as follows,

\begin{equation}
 \dot{M}_{\rm in}(r) = 2\pi r^{2} \int_{0}^{\pi} \rho \min(v_{r},0)
   \sin \theta d\theta
   \label{inflowrate}
\end{equation}
\begin{equation}
 \dot{M}_{\rm out}(r) = 2\pi r^{2} \int_{0}^{\pi} \rho \max(v_{r},0)
    \sin \theta d\theta
    \label{outflowrate}
\end{equation}
The net mass accretion rate is,
\begin{equation}
\dot{M}_{\rm acc}(r)=\dot{M}_{\rm in}(r)+\dot{M}_{\rm out}(r)
\label{netrate}\end{equation}

In a turbulent accretion flow, there are real outflows and turbulent
outflows. Real outflow means that the flows are systematically
outward moving gas. Turbulent outflow means that the outflow is not
real outflow but an outward moving portion of a turbulent eddy. In
equation 8, the outflow rate calculated include both real outflows
and turbulent outflows. Thus, the outflow rate calculated using
equation 8 is an upper limit of real outflows. The main result of
YBW2012 is that they show that systematic real outflow must be
significant and may even dominate the outflow rate calculated in eq.
8. This result is confirmed by a more recent work (Yuan et al. in
preparation). In this work, they try to follow the trajectory of
some tracer particles to judge how strong the real outflow is. They
confirm the result of YBW2012.

Usually people use the sign of Bernoulli parameter to judge whether
the outflow can escape to infinity or not.  We must point out,
however, that in a viscous or an MHD flow, the Bernoulli parameter
is not constant. Both the viscosity and magnetic field can change
the value of Bernoulli parameter during their motion. Moreover, as
shown in YBW2012,  the initial condition of the simulations is
important to determine the Bernoulli parameter of outflows. If the
Bernoulli parameter of the initial gas is negative, i.e., they are
bound, the outflow will usually be bound as well. This is the case
of Model A in YBW2012. On the other hand, if the Bernoulli parameter
of the initial gas is positive, as in the case of Model C in that
work,  the outflows are unbound (see Fig. 4 in YBW2012).

\subsection{Torus model}

In Figure 2, the top panel plots the time-averaged  (from 4-4.5
orbits) and angle-integrated mass accretion rate in model A.  The
radial profile of the inflow rate from $10 r_g$ to $200 r_g$ can be
described by
\begin{equation}
\dot{M}_{\rm in}=\dot{M}_{\rm
in}(r_{out})\left(\frac{r}{r_{out}}\right)^{0.49},
\end{equation} while it is almost constant within $10r_g$.
The initial condition and the description of viscosity in model A
are same as those in ``Run K'' of SPB99. The radial profile obtained
in SPB99 is $\dot{M}_{\rm in}\propto r^{3/4}$, steeper than ours.
The discrepancy is because the coefficient of viscosity in our model
is 10 times larger than that in ``Run K'' of SPB99 (IA99; SPB99;
Yuan, Wu \& Bu 2012).

The bottom panel of Figure 2 shows our fit to the time-averaged
inflow rate. The standard deviation of the time-averaged inflow rate
is also shown. We can see that it is not large. We have tested the
standard deviation of time-averaged physical quantities (e.g. mass
inflow rate, density, pressure and angular momentum) in other models
and found that it is not large.

Figure 3 shows the radial structure of the time-averaged (from 4-4.5
orbits) flow near the equatorial plane in  model A. The black lines
correspond to the average over angle between $\theta=84^\circ$ to
$\theta=96^\circ$. The red lines correspond to the average over
angle between $\theta=78^\circ$ to $\theta=102^\circ$. The radial
scalings of physical variables obtained over angle integral between
$\theta=84^\circ$ to $\theta=96^\circ$ are almost identical to those
obtained over angle integral between $\theta=78^\circ$ to
$\theta=102^\circ$. We conclude that it is safe to measure the
radial scalings by angle integral from $\theta=84^\circ$ to
$\theta=96^\circ$. We have done test for other models and find same
conclusion. McKinney \& Gammie (2002) also find that averaging over
$\theta=72^\circ$ to $\theta=108^\circ$ produces nearly identical
results with those averaging over $\theta=84^\circ$ to
$\theta=96^\circ$.

From the inner boundary to $200 r_g$, the density, gas pressure,
angular momentum can be described by a power law scaling with
radius,
\begin{equation} \rho \propto r^{-0.53}, p \propto r^{-1.64}, l
\propto r^{0.31}.
\end{equation} This density profile is significantly flatter than that
obtained in Yuan, Wu \& Bu (2012) where they got $\rho\propto
r^{-0.85}$. This is because in that work the initial torus is put at
a much larger radius, $R_0=10^4r_g$. Thus, the radial dynamical
range of Yuan, Wu \& Bu (2012) is much larger than the present work
and the the results suffer little from the plunging flow close to
the black hole. The density profile is similar to that obtained in
SPB99, where they found $\rho \propto r^{-0.5}$. The numerical
setting in SPB99 is different from that in this paper, SPB99 adopt
Newtonian potential and $\alpha=10^{-3}$. The former inclines to
make the density profile steeper while the latter inclines to make
it flatter. The rough consistency between SPB99 and the present work
is because the two effects cancel each other.

\begin{figure}
\includegraphics[width=8.5cm]{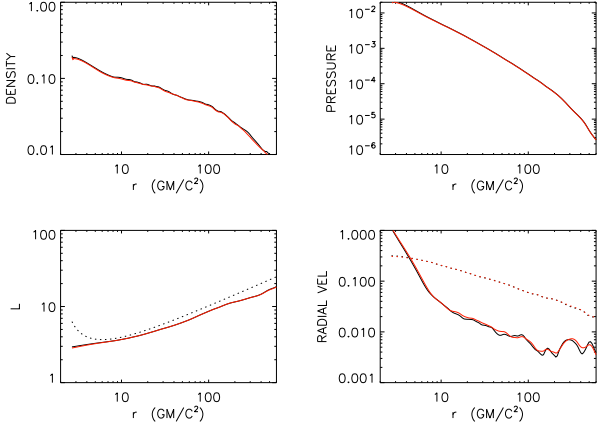}
\caption{Radial profiles of some quantities in model A. The black
lines correspond to those averaged over the polar angle between
$\theta=84^ \circ$ to $\theta=96^\circ$. The red lines correspond to
those averaged over the polar angle between $\theta=78^ \circ$ to
$\theta=102^\circ$. The dotted line in the lower-left plot
corresponds to the Keplerian angular momentum. The dotted lines in
the lower-right plot corresponds to the sound speed.}
\end{figure}

\subsection {Model B}
In model B, our initial condition is based on the one-dimensional global ADAF solution (G-ADAF-S). The radial structure of our initial
conditions is shown in Fig. 1. One feature is that the angular momentum of the
flow is smaller than that of the torus model. Fig. 4 shows the time-averaged  (from 800-900 orbits) and
angle-integrated mass accretion rate of model B.  The radial profile
of the inflow rate from $10 r_g$ to $200 r_g$ can be described by
\begin{equation}
\dot{M}_{\rm in}=\dot{M}_{\rm
in}(r_{out})\left(\frac{r}{r_{out}}\right)^{0.55}.
\end{equation} Again, the accretion rate within $10r_g$ is almost constant.
The result is approximately consistent with Model A.

Figure 5 displays the radial structure of the time-averaged (from
800-900 orbits) flow near the equatorial plane (between
$\theta=84^{\circ}$ to $\theta=96^{\circ}$). From the inner boundary
to $200 r_g$, the density, gas pressure, angular momentum can be
described by, \begin{equation} \rho \propto r^{-0.8}$, $p \propto
r^{-1.65}$, $l \propto r^{0.26}.\end{equation} We see that  the
density profile is much steeper than that in model A. We can
understand it as follows. Because of the initial condition, in the
outer region (close to $200 r_g$), the angular momentum in model A
is much higher than that of model B; therefore, the radial velocity
in model A is much smaller than that of model B. Close to the black
hole, in both models, the radial velocity is equal to light speed.
The radial profile of velocity in model A is steeper than that of
model B. Given that mass accretion rate have comparable power-law
index, the steeper radial velocity profile in model A results in
flatter density profile.

\begin{figure}
\includegraphics[width=8.5cm]{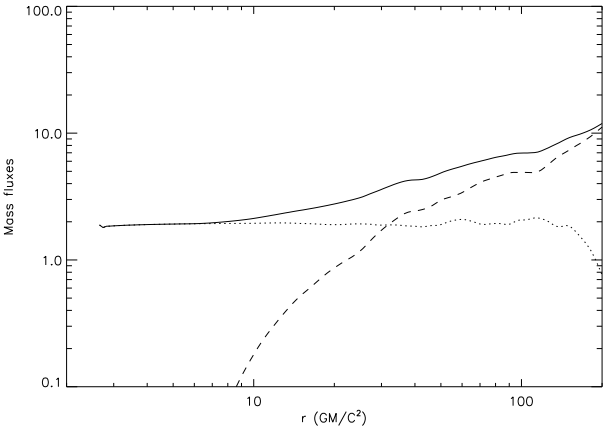}
\caption{Same with the top panel of Fig. 2, but for model B.}
\end{figure}

\begin{figure}
\includegraphics[width=8.5cm]{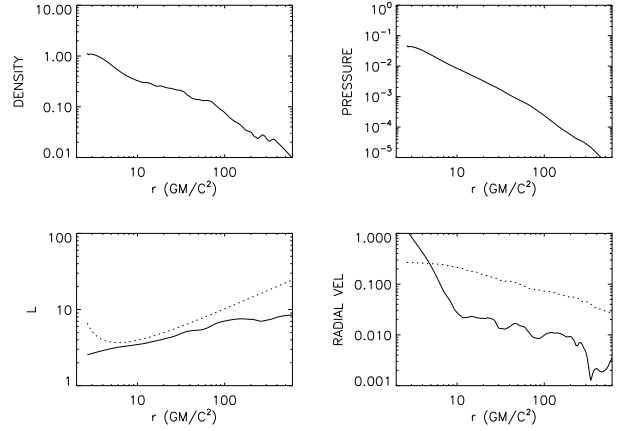}
\caption{Same with Fig. 3, but for model B. }
\end{figure}

\subsection{Injection models}
\subsubsection{Injection models with large and moderate angular momentum}
We first define the circularization radius $r_c$ as
\begin{equation}
\sqrt{\left(\frac{r_{c}}{r_{out}}\right)}=\frac{l_{inj}}{l_{ok}},
\end{equation} where $r_{\rm out}$ is our outer boundary, $l_{inj}$ and $l_{\rm ok}$ are
the angular momentum of the injected gas and the Keplerian angular momentum at $r_{\rm out}$, respectively. When $l_{inj} \ll l_{ok}$, the injected gas will quickly infall
until it get close to the circularization radius. Because the gas
pressure gradient force is important in hot accretion flow, the effective circularization radius (hereafter, ECR), where the gas stops quick infall, may be $2\sim 3$ times bigger
than  $r_c$.

Fig. 6 shows the mass accretion rates in models C1 $\sim$ C4 while
Fig. 7 shows the time averaged physical quantities near the equator
(between $\theta=84^{\circ}$ to $\theta=96^{\circ}$). In model C4,
the circularization radius $r_c \approx 37.5 r_g$, Because of the
pressure gradient force, the ECR of gas locates at $\sim 100r_g$.
Beyond ECR, we see that the angular momentum and inflow rate are
both approximately constant. The reason why the inflow rate is
constant is that convection is weak and the accretion timescale is
short (refer to \S3.3.2 for more details). In model C4, quasi-steady
state is achieved within $50 r_g$, the mass accretion rate from $10$
to $50r_g$ can be described by  \begin{equation} \dot{M}_{\rm
in}=\dot{M}_{\rm in}(r_{out})\left(\frac{r}{r_{out}}\right)^{0.54}.
\end{equation}
From the inner boundary to $50 r_g$, the density, gas
pressure, angular momentum can be described by, \begin{equation}\rho \propto r^{-0.63}, p \propto r^{-1.31}, l \propto r^{0.2}. \end{equation}

\begin{figure}
\includegraphics[width=8.5cm]{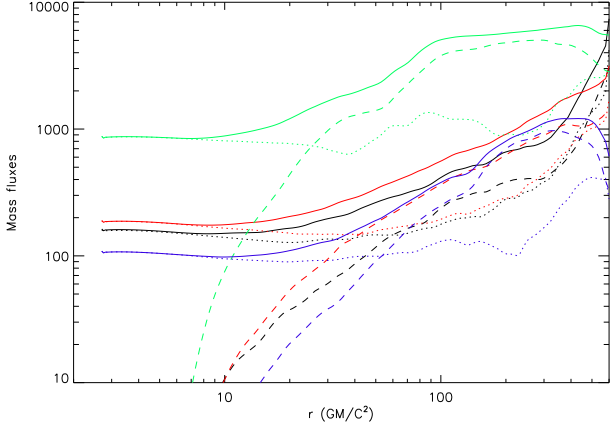}
\caption{Same with the top panel of Fig. 2, but for model C1 (black
line), C2 (red line), C3 (blue line), and C4 (green line). }
\end{figure}

\begin{figure}
\includegraphics[width=8.5cm]{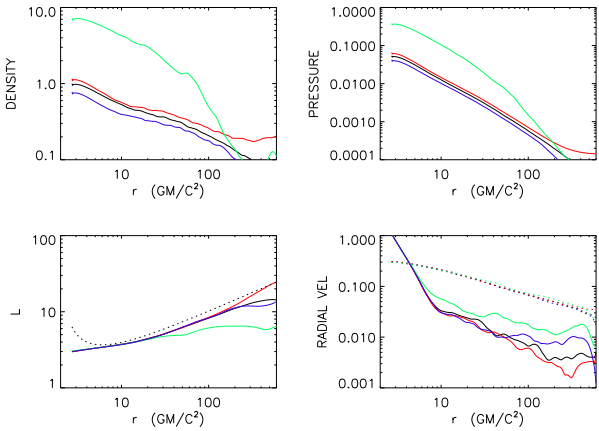}
\caption{Same with Fig. 3, but for model C1 (black line), C2
(red line), C3 (blue line) and C4 (green line). }
\end{figure}

The circularization radii of the gas in models C1 and C2 are
$541.5r_g$ and $181.5 r_g$, respectively.  The mass inflow rates of
models C1 and C2 from $10r_g$ to $200r_g$ can be described as
\begin{equation} \dot{M}_{\rm in}\propto r^{0.53}, \dot{M}_{\rm
in}\propto r^{0.54},\end{equation} respectively. The density,
pressure, and angular momentum of Model C1 and C2 can be described
by,  \begin{equation} \rho \propto r^{-0.51}, p \propto r^{-1.32}, l
\propto r^{0.31},\end{equation} and \begin{equation} \rho \propto
r^{-0.49}, p \propto r^{-1.31}, l \propto r^{0.33},\end{equation}
respectively.

\begin{figure}
\includegraphics[width=8.5cm]{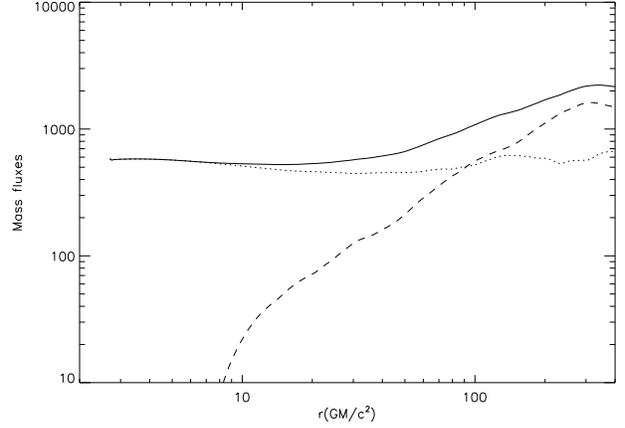}
\caption{Same with the top panel of Fig. 2, but for model D.}
\end{figure}

\begin{figure}
\includegraphics[width=8.5cm]{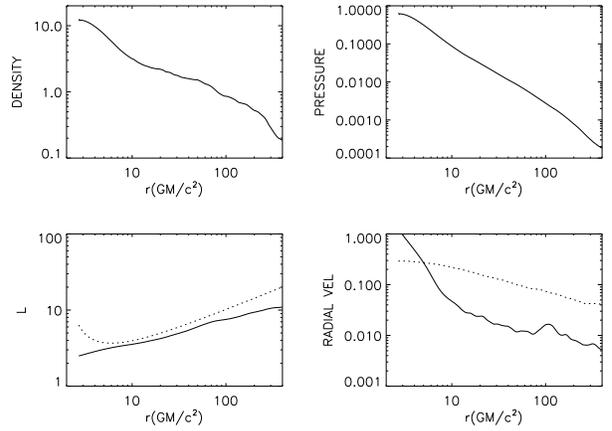}
\caption{Same with Fig. 3, but for model D.}
\end{figure}

In order to investigate the effects of injected radial velocity, we
carry out model C3. The only difference between models C2 and C3 is
that the injected radial velocity in C3 is 10 times smaller than
that in model C2. Compared to model C2, the mass accretion rate
close to the black hole in C3 is slightly smaller (by a factor
smaller than 2), the profile of inflow rate is slightly steeper,
while the slops of density, pressure and angular momentum in model
C3 almost remain same. We thus conclude that the effects of radial
velocity of the injected gas are small.

In models C1 $\sim $ C4, the properties of the injected gas are
``artificially'' given. In Model D , we adopt a ``realistic''
initial condition, i.e., the properties of the injected gas are
taken from the large-scale simulations of Sgr A* (Cuadra 2008).
Figs. 8 \& 9 show the profiles of accretion rates and physical
quantities near the equator (between $\theta=84^{\circ}$ to
$\theta=96^{\circ}$). The results are \begin{equation} \dot M_{\rm
in}\propto r^{0.47},
\end{equation}
\begin{equation} \rho \propto r^{-0.65}, p \propto r^{-1.57}, l
\propto r^{0.34}. \end{equation}

\subsubsection{Injection models with very low angular momentum}

\begin{figure}
\includegraphics[width=8.5cm]{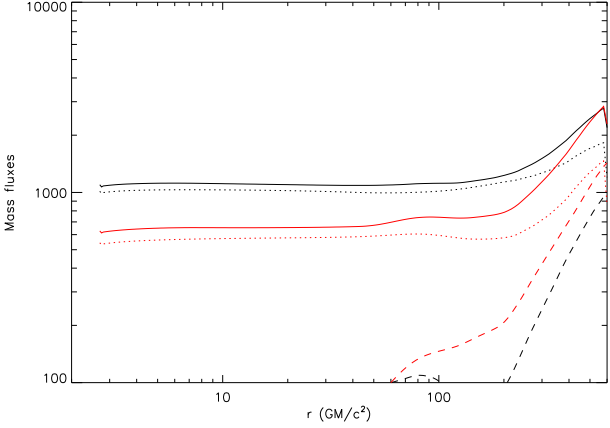}
\caption{Same with the top panel of Fig. 2, but for model C5 (black
line) and C6 (red line).}
\end{figure}

\begin{figure}
\includegraphics[width=8.5cm]{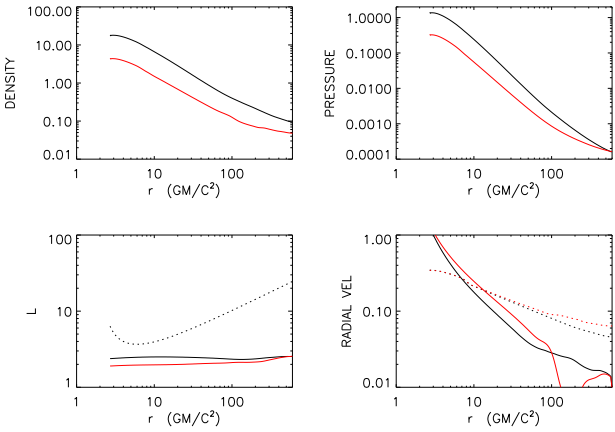}
\caption{Same with Fig. 3, but for model C5 (black line) and C6 (red line).}
\end{figure}

Previous analytical works found that when the angular momentum of
the flow is smaller than a critical value, the accretion pattern is
Bondi-like, i.e., the sonic radius is large, no matter the flow is
inviscid (Abramowicz \& Zurek 1981) or viscous (Yuan 1999). Proga \&
Begelman (2003, hereafter, PB2003) studied inviscid accretion flow
and found that when the angular momentum of the flow is smaller than
the Keplerian angular momentum at $16 r_g$, the gas inside the Bondi
radius has a constant accretion rate. Because viscosity is neglected
in PB2003, there is no heating for the gas, the entropy of the gas
is constant with radius. Therefore the flow in PB2003 is
convectively stable. When viscous heating is included, the entropy
of gas should increase inwards and convection may play some role on
the mass accretion rate profile. It is interesting then to
investigate whether in this case the mass accretion rate is
constant.

In models C5 and C6, the injected angular momentum of the gas equals
the Keplerian angular momentun at $6r_g$. In models C5 and C6, the
temperature of the injected gas is higher than Virial temperature,
so the Bondi radius $R_B=GM/c_\infty$ (where $c_\infty$ is the sound
speed at infinity) is smaller than the outer boundary, $R_B\approx
300r_g$ and $150r_g$, respectively. In reality, $R_B$ is usually
much larger.

Figures 10 and 11 show the radial profiles of mass accretion rates
and some quantities near the equator near the equator (between
$\theta=84^{\circ}$ to $\theta=96^{\circ}$) in models C5 and C6. The
results are \begin{equation} \dot{M}\approx const.
\end{equation} for both models, and correspondingly, the density
profile is close to the analytical Bondi result,\begin{equation}\rho
\propto r^ {-1.31}, \rho \propto r^ {-1.3}, \end{equation} for
models C5 and C6. The decrease of mass accretion rate close to the
outer boundary is not physical but due to boundary effects.

We analyze that the reason for a constant accretion rate is because convection is very weak.
The condition for convective instability in a rotating accretion
flow is:
\begin{equation}
N^2_{\rm eff}\equiv
N^2+\sigma^2=-\frac{1}{\rho}\frac{dP}{dR}\frac{d{\rm
ln}(P^{1/\gamma}/\rho)}{dR}+\sigma^2<0,
\label{convection}\end{equation} where $N$ is the usual
Brunt-V\"ais\"al\"a frequency and $\sigma$ is the epicyclic
frequency which is equal to rotation angular velocity $\Omega$ for
nearly Keplerian rotation (Narayan \& Yi 1994). For a non-rotating
flow, $\sigma=0$, this condition is then equivalent to an inward
increase of entropy, which is the well-known Schwarzschild
criterion. In the accretion process, the inward increase of entropy
is because of viscous heating. The viscous heating rate is
proportional to square of gradient of angular velocity. In models C5
and C6, the gradient of angular velocity is very small. Therefore,
viscous heating rate is very small in models C5 and C6, which
results in a small entropy gradient in models C5 and C6. The first
term in equation (\ref{convection}) is proportional to gradient of
entropy. We find in almost half computational domain the flow is
convectively stable according to equation (\ref{convection}). In
addition, the accretion timescale of the flow in models C5 and C6 is
very small. We find that only in a very small region around
$100r_g$, the growth timescale of convection is slightly smaller
than the gas accretion timescale. This is why convection is
unimportant\footnote{In the MHD simulation of Bondi accretion flow,
under some initial configuration of the magnetic field, Igumenshchev
\& Narayan (2002) find significant magnetic reconnection heating
thus entropy production. In this case, the flow is convectively
unstable and $\rho\propto r^{-0.5}$.}.

\subsection{Effects of boundary conditions}

In numerical simulations, for the radial boundaries, the outflow
boundary conditions are usually adopted. This condition means that
the hydrodynamical variables (e.g. density, velocity, pressure) are
directly copied from the first (or last) active zone to the ghost
zone. In spherical coordinate, the mass flux is expressed as $r^2
\rho v_r$. If we use outflow boundary conditions, the gradient of
mass flux is not zero at radial boundaries, thus there may be
artificial mass accumulation. In order to avoid this problem, some
authors use mass-flux conservation boundary condition which assumes
that the gradient of mass flux cross the boundaries is zero.

We have studied how these two boundary conditions affect the
properties of accretion flow in the torus model. For the inner
radial boundary, we find that the properties of the flows are very
insensitive to the inner boundary conditions. The reason is that at
the inner boundary, the flow is supersonic, so no wave signal can
transfer back to affect the properties of the flow (McKinney \&
Gammie 2002). In order to study how the two boundary conditions at
the outer radial boundary affect the properties of the flow, we
carry out model A1. The only difference between models A1 and A is
the outer radial boundary conditions. In models A1 and A, we use
mass-flux conservation boundary condition and outflow boundary
condition, respectively. Fig. 12 plots the time-averaged (from 4-4.5
orbits) and angle integrated mass accretion rate in the two models.
Fig. 13 shows the radial structure of the time-averaged flow near
the equatorial plane ($\theta=84^o$ to $\theta=96^o$) in the two
models. From Fig. 12 and Fig. 13, we find that the properties of the
flow are very insensitive to the outer boundary conditions.
\begin{figure}
\includegraphics[width=8.5cm]{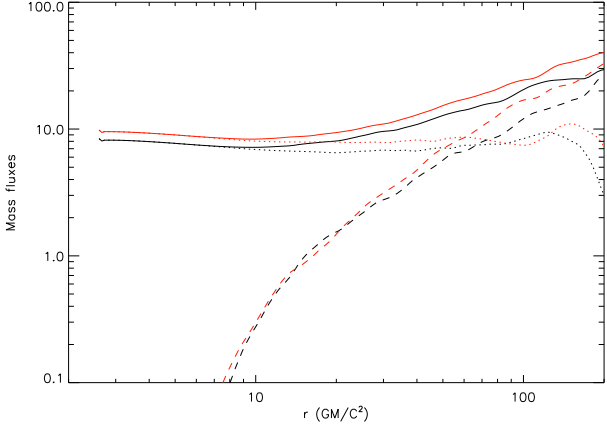}
\caption{Same with the top panel of Fig. 2, but for models A (black
lines) and A1 (red lines).}
\end{figure}

\begin{figure}
\includegraphics[width=8.5cm]{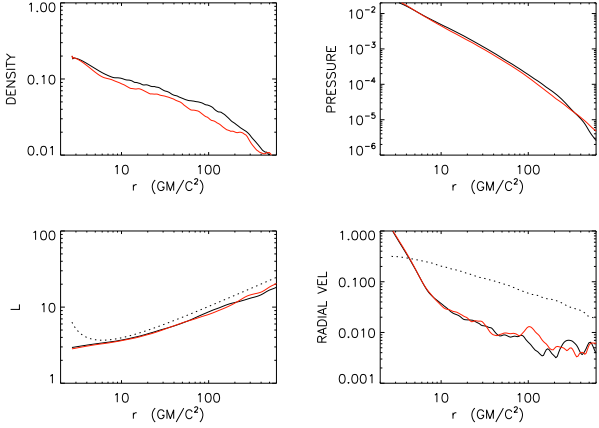}
\caption{Same with Fig. 3, but for models A (black lines) and A1 (red
lines).}
\end{figure}

\subsection{Dependence on resolution}
In order to test whether the results obtained in this paper depend
on resolution, we carry out models Ah1 and Ah2. For model Ah1, the
numbers of grids in $r$ and $\theta$ directions are 1.5 times those
in model A. In model Ah2, the resolution is 2 times higher than that
in model A.

Fig. 14 plots the time-averaged (from 4-4.5 orbits) and angle
integrated mass accretion rate in models A (black lines) and Ah1
(red lines) and Ah2 (green lines). From this figure, we can see that
the power-law indexes of the inflow rate in torus models with
different resolution are almost same. Fig. 15 shows the radial
structure of the time-averaged density near the equatorial plane
($\theta=84^o$ to $\theta=96^o$) in models A (black line) and Ah1
(red line) and Ah2 (green line). The radial scaling of density with
different resolution is also almost same. From Fig. 14 and Fig. 15,
we find that the resolution used in most of the models ($168 \times
88$) in this paper is enough. 
\begin{figure}
\includegraphics[width=8.5cm]{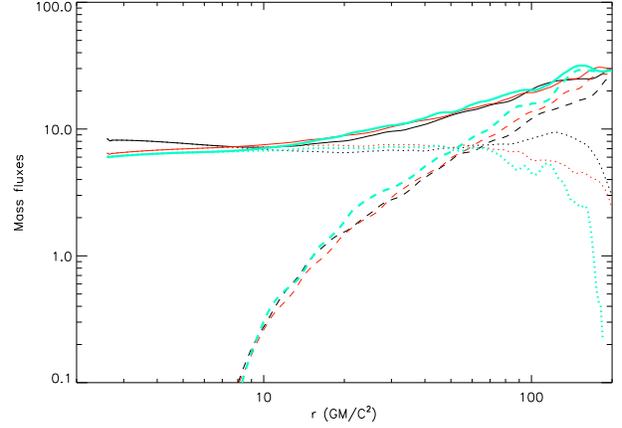}
\caption{Same with the top panel of Fig. 2, but for model A (black
lines) and Ah1 (red lines) and Ah2 (green lines).}
\end{figure}

\begin{figure}
\includegraphics[width=8.5cm]{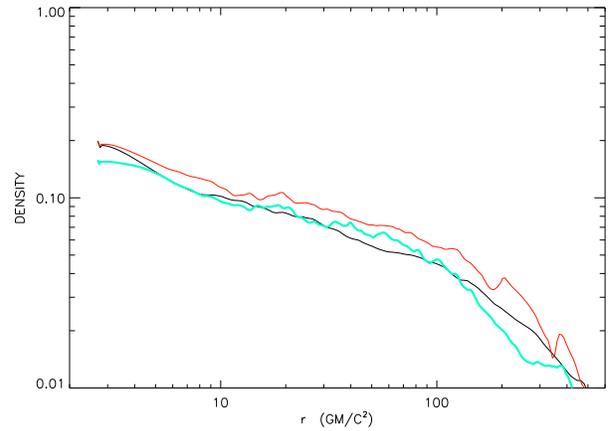}
\caption{Radial profiles of density in models A (black line) and Ah1
(red line) and Ah2 (green line). The quantity has been averaged over
time and the polar angle between $\theta=84^ \circ$ to
$\theta=96^\circ$.}
\end{figure}

\section{SUMMARY}

In this paper, we have investigated the effects of initial and
boundary conditions on the properties of hot accretion flow using
two-dimensional hydrodynamic simulations.  We have considered
several initial conditions, including the most widely adopted
``torus'' and ``injection'' conditions, and the more realistic flow
properties from large-scale simulation. Special attention is paid to
the radial profiles of mass accretion rate and density. Both
profiles can be described by a power-law function ($\dot{M}_{\rm
in}\propto r^{s}, \rho\propto r^{-p}$). The simple description to
different initial conditions and the corresponding power-law index
are summarized in Table 1.

The most interesting result is that for various initial conditions,
if the angular momentum of the accretion flow is not too low so that
a rotationally supported accretion flow can form beyond $10r_g$, the
power-law index of the radial profile of inflow rate $s$ lies in a
narrow range of $0.47\la s \la 0.55$. But on the other hand, the
radial profile of density is more sensitive to the initial
condition. The power-law index $p$ lies in a wider range $0.48\la p
\la 0.8$. Given that the inflow rate profile is almost  same for
various initial conditions, the diversity of the density profile is
because different initial conditions give different angular momentum
profiles and thus different radial velocity profiles. Then it is an
interesting question  why different initial conditions give roughly
same inflow rate profile. The readers are referred to Begelman
(2012) for an explanation.

Since $\dot{M}\propto r^2 v_r \rho(r)$, for hot accretion flows, we
usually expect $p=1.5-s$, if $v_r \propto r^{-0.5}$ according to the
self-similar scaling law of hot accretion flows (e.g. Narayan \& Yi
1994). We note that our results do not satisfy $p=1.5-s$, but
smaller than this value. This is partly because the density profile
depends not only on the inflow rate profile, but also on outflow
rate profile, which is much steeper. Another reason is that
$v_r\propto r^{-0.5}$ is not well satisfied since the flow is close
to the black hole.

If the angular momentum of the flow is very low (i.e., models C5 and
C6), we find that inside the Bondi radius, the mass accretion rate
is almost constant. Physically, this is because convection is very
weak, thus, very little outflow is produced. In these models, the
viscous heating rate is very small due to the small angular velocity
gradient. Small viscous heating rate results in a very small
gradient of entropy, thus, convection is weak. In addition, the
accretion timescale is very small.

We have also studied the effects of boundary conditions. We have
considered two widely adopted inner and outer boundary conditions,
which are ``outflow'' boundary condition and ``mass flux
conservation'' boundary condition. We have found that the properties
of the flow are very insensitive to the radial boundary conditions.

\section{ACKNOWLEDGMENTS}

This work was supported in part by the Natural Science Foundation of
China (grants 11103059, 11121062, and 11133005), the National Basic
Research Program of China (973 Program 2009CB824800), and the
CAS/SAFEA International Partnership Program for Creative Research
Teams. J. C. acknowledges support from FONDECYT (11100240) and Basal
(PFB0609). The simulations were carried out at Shanghai
Supercomputer Center.

\end{document}